# Gold Nanoparticles for Plasmonic Biosensing: The Role of Metal Crystallinity and Nanoscale Roughness*

Jean-Claude Tinguely**[1], Idrissa Sow[2], Claude Leiner[1], Johan Grand[2], Andreas Hohenau[1], Nordin Felidj[2], Jean Aubard[2], Joachim R. Krenn[1]

[1] *Institute of Physics, Karl-Franzens University, Universitätsplatz 5, A-8010 Graz, Austria*

[2] *Laboratoire ITODYS, Université Paris Diderot, CNRS UMR 7086, 15 rue Jean de Baïf, F-750013 Paris, France*

Noble metal nanoparticles show specific optical properties due to the excitation of localized surface plasmons that make them attractive candidates for highly sensitive bionanosensors. The underlying physical principle is either an analyte-induced modification of the dielectric properties of the medium surrounding the nanoparticle or an increase of the excitation and emission rates of an optically active analyte by the resonantly enhanced plasmon field. Either way, besides the nanoparticle geometry the dielectric properties of the metal and nanoscale surface roughness play an important role for the sensing performance. As the underlying principles are however not yet well understood, we aim here at an improved understanding by analyzing the optical characteristics of lithographically fabricated nanoparticles with different crystallinity and roughness parameters. We vary these parameters by thermal annealing and apply a thin gold film as a model system to retrieve modifications in the dielectric function. We investigate, on one hand, extinction spectra that reflect the far-field properties of the plasmonic excitation and, on the other hand, surface-enhanced Raman spectra that serve as a near-field probe. Our results provide improved insight into localized surface plasmons and their application in bionanosensing.

*Keywords: localized surface plasmons, Surface-enhanced Raman scattering, roughness, biosensors, nanosensors*

# 1. Introduction

Nanoparticles made of metals as gold, silver or copper differ strongly from the respective bulk material in their optical response due to the excitation of localized surface plasmons (LSPs). These coherent oscillations of the conduction electrons upon excitation by light give rise to a strong extinction band in the visible or near infrared-spectral range. The according spectral position is highly sensitive to the used metal, size, shape and dielectric surrounding [1]. Responding sensitively to changes in the surrounding medium as well as generating resonantly enhanced nanoscale light fields, LSPs can be applied for simple, real-time and label-free biosensing schemes [2]. In particular, when binding molecules to the surface of LSP-particles, the according increase in the refractive index causes the extinction resonances to red-shift, as reported in various studies on the real-time molecular binding monitored by simple optical transmission spectrometry [3-5]. Among the studied analytes, LSP-based detection has been essayed with, e.g., heavy metal ions [6], toxin [7], glucose [8], nucleic acids [9], biotin-streptavidin [10], or antigen-antibody interactions [11,12]. An ordered array of similar particles is usually applied for a high signal-to-noise ratio, with recent developments aiming at single molecule detection [13,14]. The resonantly enhanced plasmonic near-fields also offer the possibility of coupling LSP with photon emitters such as quantum dots or molecules to exploit molecular identification techniques as surface-enhanced fluorescence (SEF) [15] or surface-enhanced Raman scattering (SERS) [16,17]. For the latter, signal enhancements of over $10^{10}$ have been reported [14].

Gold is frequently favored as a chemically inert plasmonic material. Fabrication methods for metal nanoparticles include nanosphere lithography (NL) [18], electron beam lithography (EBL) [19] and the chemical synthesis in liquid media [20]. As lithography methods, NL and EBL employ a masking structure to first define the geometry of the nanoparticles that are then grown by, e.g., vacuum deposition of the metal. This process results in a polycrystalline metal structure, in contrast to the mostly single-crystalline particles fabricated by chemical synthesis. When highlighting the different process advantages, NL stands out for its simplicity and broad area coverage, EBL for the freely selectable shape of the nanostructures and chemical synthesis for the well-defined particle crystallinity. As disadvantages, there are shape restrictions for NL, low throughput and high instrumental costs for EBL and, compared to the first two methods, a challenging ordered deposition on substrates and size inhomogeneities for chemical synthesis. Generally speaking, EBL is arguably the most flexible method for nanoparticle fabrication. It can be combined with some control over the metal crystallinity parameters through thermal curing ("annealing"). The influence of the crystalline structure and surface roughness of metal nanoparticles on their optical properties remains however to some extent a motive of speculation, while its drastic impact was made clear in some cases by the comparison between chemically synthesized and EBL-fabricated structures [21]. On the one hand it can be argued that roughness and polycrystallinity give rise to a stronger damping of LSP modes due to increased electron scattering rates [22-24]. This effect weakens the LSP resonance and is thus disadvantageous for sensing. On the other hand, however, roughness features can give rise to locally strongly enhanced optical near-fields dubbed "hot spots" that are considered to be of primary importance for surface-enhanced effects as SERS and SEF [14,25]. Although these aspects present important implications for the development of LSP sensors, only few studies have been dedicated to the investigation of the influence of crystallinity and roughness on the plasmon response and near-field enhancement of metal nanoparticles. In [26], Huan et al. concluded a decrease of the electron-phonon relaxation rate in NL-fabricated gold nanoprisms due to strong thermal annealing. Chen et al. investigated square-shaped nanoparticles, relating a decrease of the Drude relaxation rate to an annealing-induced reduction of internal defects [26]. In [27], Rodriguez-Fernandez et al. concluded that the LSP fields depend sensitively on roughness for spherical gold colloids. Modeling a symmetrical surface corrugation led to the similar conclusions in [28]. However, Trügler et al. came to a contrary finding by means of single particle spectroscopy of EBL samples and the modeling of random fluctuations of surface roughness [29], namely a weak dependence of the far-field response of plasmonic nanoparticles on details of the surface morphology.

This paper aims to discuss the respective influence of the crystalline structure and nanometric surface roughness for EBL-fabricated LSP-sustaining gold nanoparticles. Therefore, the metal structure is modified through thermal annealing. In a first step, ellipsometry measurements of a continuous gold thin film together with surface imaging by scanning electron microscopy (SEM) and atomic force microscopy (AFM) will aid to track the impact of the thermal treatment. In particular, we monitor modifications in the gold crystal structure via the dielectric function deduced from the ellipsometry measurements, while AFM provides data on the surface roughness. Then, optical extinction spectra and SERS measurements are applied to identify the annealing-induced changes on the LSP properties, and conclusions with respect to the biosensing performance are drawn.

# 2. Gold Thin Films

As the size of single crystallites (some to tens of nanometer) within EBL-fabricated gold nanoparticles is usually smaller than the nanoparticle diameter (around 100 nm), extended thin gold films can be considered as a model system for evaluating annealing-induced modifications in dielectric function and surface roughness.

## 2.1 Growth and Crystallisation of Gold Thin Films

A simple thermodynamic description can be used to elucidate the gold film growth process as applied in EBL [30]. At deposition, the gold atoms adsorbed at the surface may wander freely and bind to other adsorbed atoms to form clusters (nucleation process). The clusters will grow as the probability of an atom to get adsorbed exceeds their desorption rate. By binding to the surface, free energy is lost leading to a more stable state, with surface imperfections as favored positions for growth. For the growth of metal layers on glass substrates, the crystallite orientation is a function of the respective surface tensions at the interfaces crystallite-substrate, crystallite-gold vapor, and substrate-gold vapor, all related through the wetting angle using Young´s equation. With metals yielding a much higher surface tension than the dielectric interface, a positive wetting angle value in Young's equation is obtained that explains the island growth behavior of gold films below ca. 20nm mass thickness [30]. As the clusters expand and touch each other, the layer growth begins. Due to the different orientations developed during the island growth phase, the boundaries between the clusters are expected to be maintained resulting in columnar-like crystallites [31].

## 2.2 Annealing Effects on Gold Thin Films

Thermal annealing is employed in thin film applications for a long time for reducing strain and the number of intrinsic or extrinsic defects in crystal lattices. Studies of the grain evolution of thin gold films submitted to heat treatment sketch the following model [23,25]. Upon a 200°C exposure the average grain size increases while surface roughness decreases, as intuitively expected. Grain growth around this temperature is expected to occur mainly through grain boundary wandering towards the particle borders instead of direct coalescence of neighboring grains. When approaching equilibrium during annealing, small intersection angles ($\alpha$, inset Fig. 1a) caused by high surface energies relax towards a maximal value ($\alpha_{max}$, inset Fig. 1b), determined by the corresponding interfacial energies. Experiments also showed that the surface morphology adapts to the grain boundary movement, and concerning crystallinity, only (111) orientation is usually detectable after annealing, being the most stable for gold as an fcc-metal [23]. Stress relaxation combined with surface diffusion may therefore result in broader gaps, leading to a more prominent contribution of the crystallographic features to the surface morphology as compared to the impression of topographical effects (Fig. 1).

In Fig. 1, SEM images of a 40 nm thick gold thin film on quartz substrate (thermal vacuum deposition at room temperature, background pressure $3 - 8 \times 10^{-6}$ mbar, rate 0.3 – 0.8nm/s for all samples in this work) are compared, before and after annealing on a 200°C hot plate for 20min. The grained structure before the annealing process correlates well with the grain size of ca. 30nm estimated through various techniques as X-ray diffraction [32], AFM [33] and transmission electron microscopy (TEM) [23,33]. As the average lateral size of around 100nm for the apparent grains after annealing again corresponds to values deduced from TEM measurements [23], we conclude that the SEM data reflect the actual grain size. From complementary AFM images (found in the Supporting Information) we deduce a surface roughness of 0.9 ± 0.02nm rms for the as-evaporated sample and 0.7 ± 0.02nm rms for the sample after annealing.

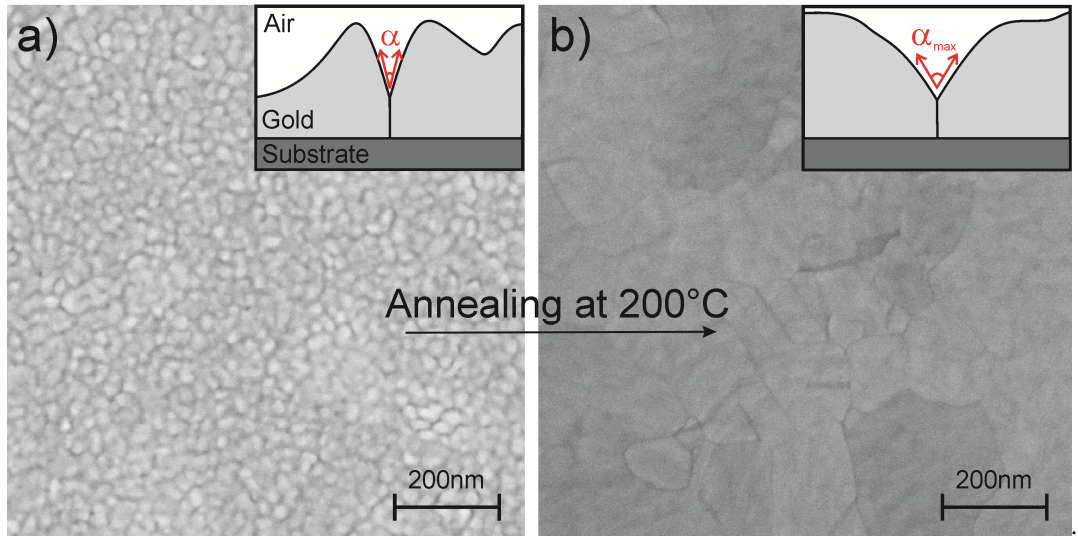

**Fig. 1** SEM images of a 40nm thick gold film (a) before and (b) after annealing at 200°C for 20min. The insets sketch the grain boundary interfaces, broader gaps resulting from stress relaxation and surface diffusion (horizontal and vertical axes of insets not to scale)

According to studies in the Drude and interband transition regions of the dielectric function [24], certain aspects can be attributed to the annealing-induced changes. A crystallite size of ca. 30nm is comparable to the mean free path of the electrons in gold. Annealing the sample and increasing the average crystallite size to ca. 100nm will thus result in less electron scattering events and less damping. Additionally, the observed smoothening in the surface could lead to reduced damping as well, since surface roughness features act potentially as electron scatter centers. Indeed, ellipsometry measurements (Supporting Information) on the thin film indicate the validity of this model, as a modified dielectric function with decreased values of both, real, and imaginary part are found for the annealed film.

## 3. Gold Nanoparticles

We now turn to the investigation of gold nanoparticles and the impact of annealing on the LSP resonance by extinction spectroscopy. Furthermore, SERS measurements will be discussed to illustrate according consequences for biosensing.

## 3.1 Fabrication and Annealing of Gold Nanoparticles

The gold nanoparticles investigated in the following are fabricated by EBL on quartz substrates. To produce the samples, 2% poly (methyl methacrylate) solution (Allresist Chemicals) is spin casted on the quartz surface rendering a ca. 100nm positive resist layer. The resist is then patterned by exposure by an electron beam at designed positions in a computer-controlled electron microscope (Raith 100-2; acceleration voltage, 20 kV; electron dose, 300 µC/cm2; electron current, 11pA). By wet-chemical development (AR600-55, Allresist Chemicals) for 40seconds, the exposed areas are removed leaving a mask for the subsequent thermal vacuum deposition of gold (40nm, as described in Section 2.2). Immersing the sample in 45°C acetone for two hours removes the remaining PMMA and the gold at the non-exposed regions. Here, the disk-shaped particles remaining at the substrate have nominal diameters of 150, 175, 200, and 225nm, a height of 40nm and are arranged in a square pattern (300nm lattice constant) over array areas of 100x100µm$^2$. The SEM images in Fig. 2 (a) and (b) depict the same six nanoparticles with a (nominally identical) diameter of 225nm before and after annealing at 200°C for five minutes. The annealing parameters were chosen to avoid any significant modification of the particle geometry (see below).

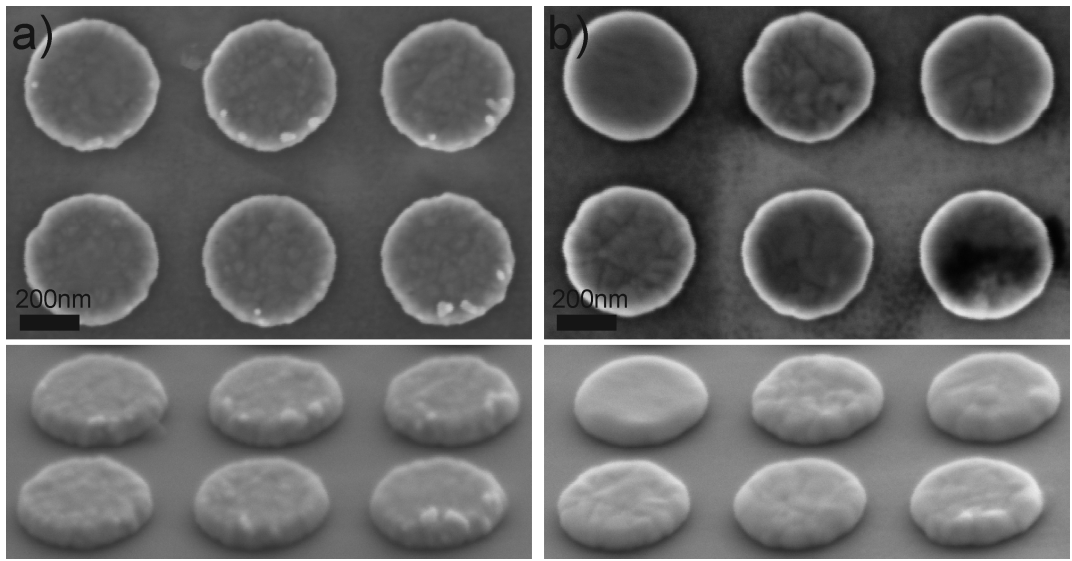

**Fig. 2** SEM images of gold nanoparticles (a) as-evaporated and (b) annealed at 200° for five minutes. The top panels show the samples in top view, the bottom panels at an observation angle of 60°. In contrast to the samples used for the following measurements this sample includes a conductive indium tin oxide layer on top of the substrate to prevent charging and thus allowing detailed SEM imaging

By comparing the images before and after annealing, several observations can be made. First, there is a clear reduction in surface roughness due to annealing, e.g., the spurious features close to the particle circumference presumably originating from the lift-off process are smoothened. Second, the crystalline grains have increased in size. While a clear assignment of individual grains before and after annealing is not possible from the SEM images, it seems that individual particles respond quite differently to the heat treatment. For example, while most particles appear polycrystalline and somewhat rough, the nanoparticle in the upper left appears exceptionally smooth after annealing without any indication of grain boundaries. The average nanoparticle dimensions as deduced from the SEM images in Fig. 3 are 229 ± 4 and 226 ± 4 nm before and after the annealing, respectively. The reduction in lateral size is most probably due to surface tension. This conclusion is backed by complimentary AFM images (Supporting Information) indicating a minor increase of the nanoparticle height. This does not, however, present a significant modification of the nanoparticle geometry, leading us to conclude that the particle shape (an important parameter setting the LSP resonance wavelength) is not significantly modified upon the annealing conditions chosen here. Particle surface roughness was estimated from AFM measurements on 49 particles with nominal diameters of 225nm, resulting in a mean value of 2.6±1.5nm rms before and 1.4±0.5nm rms after annealing (AFM images and the surface roughness estimation procedure can be found in the Supporting Information).

### 3.2 .Extinction Measurements

To quantify the optical response of the nanoparticle arrays as discussed in the previous section, we now turn to probing the optical far-field of the LSP resonances by measuring extinction spectra with a microspectrograph based on an optical microscope with a fiber-coupled spectrometer. The curves in Fig. 3 (a) show the spectra for four arrays containing particles of different diameters, both as-fabricated (dashed lines) and annealed at 200°C for 5 minutes (full lines). We find that in all cases the annealing process induces a blue-shift of the LSP resonance wavelength of ca. 20nm. We note that for the chosen parameters the resonance position of an annealed particle array (e.g., diameter 225 nm) coincides well with the resonance position with an array of non-annealed smaller particles (e.g., diameter 200 nm). Furthermore, we observe that annealing leads to a reduction of the full-width-at-half-maximum (FWHM) of the spectra as compared to the non-annealed samples, and to accordingly higher extinction.

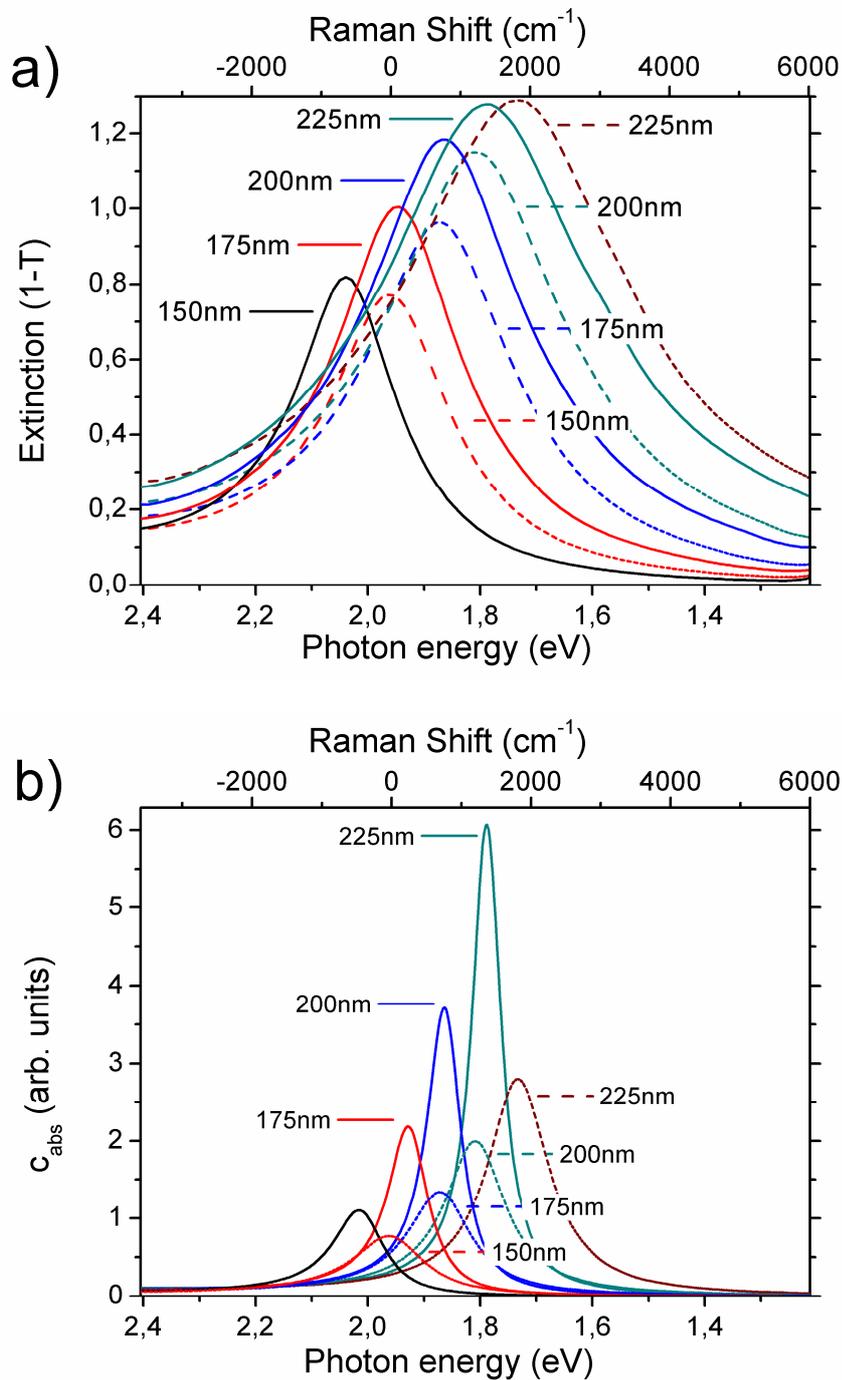

Fig. 3 (a) Experimental extinction spectra of arrays of gold nanoparticles on quartz substrate with 150, 175, 200, and 225nm diameter, before (dashed lines) and after (solid lines) annealing at 200°C for five minutes. (b) Quasistatic calculation of oblate nanoparticles corresponding to (a). The upper axes are depicted as a Raman shift scale for later comparison with SERS spectra, 0 cm$^{-1}$ corresponding to the SERS excitation wavelength of 633nm. The scale conversion is achieved by subtracting the photon energy values from the excitation energy, both in cm$^{-1}$ units.

We now show that the observed annealing-induced changes can be attributed largely to changes in the dielectric function of gold. Therefore, we plot in Fig. 3 (b) the results of calculations relying on a simple quasistatic model [34-36], for simplicity assuming particles in a homogeneous dielectric environment. We calculate the particle absorption as $A \propto \pi \cdot \lambda^{-1} \cdot \mathrm{Im}(\alpha)$, where $\lambda$ is the wavelength and $\alpha$ the particle polarizability given by $\alpha = V \cdot (\varepsilon_m - \varepsilon_d) \cdot (\varepsilon_d + L(\varepsilon_m - \varepsilon_d))^{-1}$ (V is the particle volume and $\varepsilon_m$, $\varepsilon_d = n_d^2$ are the dielectric functions of the metal and the homogeneous dielectric environment). The shape factor *L* was fitted to give the experimentally observed spectral resonance positions for an effective refractive index of the dielectric environment of $n_d = 1.27$ [1] (Fig. 3(b), dashed curves). With the same shape factors, the calculations were then repeated using the dielectric function measured by ellipsometry from a gold film annealed at 200°C for 5min (Supporting Information). The experimentally observed qualitative trend, a blue shift, a narrowing, and an increase of the LSP peak is nicely reproduced and, at least for the larger particles, the values of the spectral shift correspond well to the measured values (full curves in Fig. 3 (b)). The accordance between measured and calculated values demonstrates that particle geometry changes have no dominant effect on the spectra. We note that the much lower spectral width in the calculated data as compared with the measurements can be qualitatively attributed to the increased LSP damping in the sample due to radiation damping and particle-particle near-field interactions in the closely spaced nanoparticle arrays, effects that are not included in the quasistatic model.

We note that surface roughness was recently shown to have only a minor impact on the far-field properties of individual nanoparticles [29] . Therefore, scattering spectra of single EBL-fabricated particles were acquired through a dark-field configuration. Although differing in individual nanoscale roughness, plasmon resonance positions, and spectral widths of the particles show only minor variations. Numerical simulations adopting the shapes of the investigated particles as extracted from SEM images supported these results. In addition, a further analysis was carried out through a quasistatic analysis of cylindrical nanorods adding stochastic height variations to the surface. For calculated spectra with different height-to-diameter ratios averaged over 100 random realizations of surface roughness, it was again shown that the spectral signature of the particles does not change, apart from a small red shift and a minor individual spectral variance for individual particles with increasing roughness height difference.

---

[1] The refractive index of 1.27 was chosen as a weighted mean of quartz (n=1.46 at 700nm, 60%) and air (n=1, 40%) assuming a slightly stronger impact of the higher refractive medium. The spectral shift induced by using the dielectric functions of the annealed compared to the non-annealed gold films is largely independent of this choice.

In addition, we checked with AFM and SEM the uniformity of the individual particles within an array, as shape deviations could lead to inhomogeneous spectral broadening. We found the particles to be rather uniform so that we can exclude such a broadening effect. We also conclude that the annealing leads to larger grains and accordingly less grain boundaries in the nanoparticles, as also observed for the thin film. The reduced damping corresponding to the reduced dielectric function as measured from the gold thin film is nicely mirrored in the spectrally narrower LSP peak, clearly showing an improved quality factor of the resonant LSP mode. This, in turn, corresponds to a stronger LSP resonance and thus to a higher optical near-field around the nanoparticle as compared to the non-annealed sample. This should be advantageous for LSP-based sensing applications.

### 3.3. Surface-Enhanced Raman Spectroscopy

Having found strong indications of enhancing the LSP near-field upon annealing of nanoparticles, we now turn to investigating the consequences for biosensing performance by SERS measurements on the nanoparticle samples. Generally speaking, the SERS intensity depends strongly on the local electric field experienced by the molecules which should thus increase for LSP modes that are less damped. Very high local fields can however be generated by surface features with high curvature corresponding to roughness features ("hot spots") [14] . Our measurements shall thus allow identifying if the primary SERS source is due to the average LSP field (increasing field strength upon annealing due to dielectric function modifications) or "hot spots" (decreasing field strength upon annealing due to decreasing roughness).

For the measurements, we deposit Rhodamine 6G molecules by immersing the nanoparticle arrays for 10 minutes in a $10^{-6}$ molar aqueous solution, followed by water rinsing and $N_2$ drying. The spectra plotted in Fig. 4 were measured with a LabRam HR (Horiba Jobin Yvon), using a 100x objective in backscattering geometry and a helium-neon laser to excite the particle arrays at a wavelength of 633nm. Let us first discuss the difference of the SERS signal observed for a fixed type of particle array before and after annealing, e.g., the array with the 175nm particles (Fig. 4). It is important to distinguish between the fluorescence background [37] and the specific SERS intensity signals (measured as the background-subtracted area under one SERS peak, see the shaded area in Fig. 4 for an example). It has to be considered, that the (specific) SERS intensity as well as the SERS background depend on the spectral position of the LSP resonance with respect to the laser excitation and the spectral shifts upon annealing. The largest specific SERS gains are obtained, if the spectral LSP resonance position is centered between the laser excitation and a Raman line [38]. As the LSP resonance shifts to the blue upon annealing, we accordingly expect an increase (decrease) in the specific SERS signals for the array with the 200nm (175nm) sized particles.

We observe in all cases a specific SERS intensity on the annealed arrays that is lower than on the non-annealed sample, although the LSP extinction spectra demonstrate that quality factor and extinction are increased. We note that the SERS background changes its spectral characteristic [39] and is partly increased (e.g., 175nm samples) or decreased (150nm samples). For the sake of comparability, the particle dimensions in our samples were selected in a way that the spectral LSP position of one array before annealing is (nearly) identical to the LSP position of the array of the next larger particles after annealing (Fig. 3 (a)). By comparing the SERS intensities of the pairs of arrays with analog resonance position (Fig. 4 (a), e.g., 150nm, non-annealed with Fig. 4(b), 175nm annealed etc.), one finds that both, specific SERS intensity and SERS background signal are smaller after annealing, even for samples which should benefit of a maximal Raman gain as discussed above.

These observations allow us to draw the conclusion that the crystallite growth and smoothening of the particles has a negative effect on the specific SERS enhancement of the array, although the LSP resonance (as observed in the extinction spectra) shows increased amplitude and less damping due to improved dielectric properties. The argumentation is further substantiated by the fact that the comparison comprehends non-annealed, smaller particle arrays put against annealed, larger particles presenting higher surface area for molecule binding. These aspects point towards a considerable contribution of the yet hardly controllable nanoscopic surface roughness and crystallinity to the SERS efficiency of lithographically fabricated substrates.

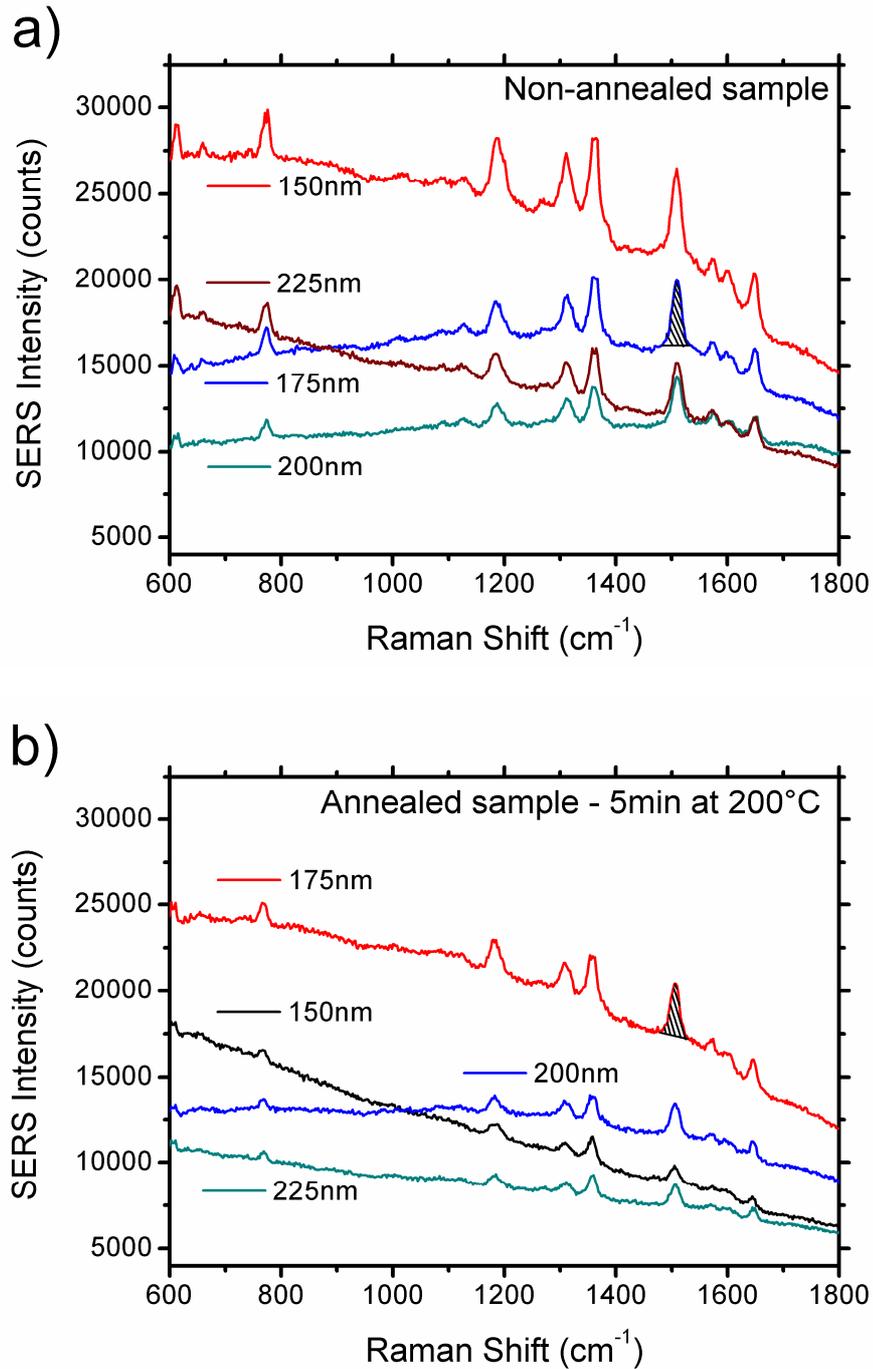

**Fig. 4** SERS spectra of rhodamine-6G for an excitation wavelength of 633nm acquired from (a) non-annealed and (b) annealed nanoparticle arrays. The different nanoparticle diameters are given in the images, identical colors in (a) and (b) correspond to practically identical LSP resonance frequencies. The shaded areas indicate the specific SERS intensity used for comparison between different spectra

# 4. Conclusions

Although improving the LSP resonance quality factor through thermal annealing, as verified by extinction spectroscopy and quasistatic calculations, our results show that the efficiency of near-field processes as SERS is reduced, presumably due to changes in surface morphology. Great care has thus to be taken when using extinction spectra as a reference for deducing near-field effects. For the sake of completeness, we note that in principle, the annealing process could alter the molecular adsorption characteristics on the particles. While there is no indication yet of such an effect, this aspect deserves further investigation.

Finally, this study contrasts the good reproducibility of far-field properties against the challenging reproducibility of near-field properties due to hardly controllable crystallization behavior. This is of particular impact for electron beam-lithographed SERS substrate fabrication, as recrystallization also takes place at room temperature, although on much lower rates.

Acknowledgements

The authors would like to thank Georg Jakopic of the Joanneum Research Institute in Weiz for the ellipsometry measurements. This work has been supported in part by the Austrian Fonds zur Förderung der wissenschaftlichen Förderung (FWF) under project no. 21235-N20.

References

[1] Maier SA (2007) Plasmonics - fundamentals and applications. Springer, New York.

[2] Anker JN, Hall WP, Lyandres O, Shah NC, Zhao J and Van Duyne RP (2008) Biosensing with plasmonic nanosensors. Nat. Mater. 7:442-453. DOI:10.1038/nmat2162 ER.

[3] Willets KA and Van Duyne RP (2007) Localized surface plasmon resonance spectroscopy and sensing. Annu. Rev. Phys. Chem. 58:267-297. doi: 10.1146/annurev.physchem.58.032806.104607.

[4] Yonzon CR, Stuart DA, Zhang XY, McFarland AD, Haynes CL and Van Duyne RP (2005) Towards advanced chemical and biological nanosensors - an overview. Talanta 67:438-448. DOI:10.1016/j.talanta.2005.06.039 ER.

[5] Dahlin AB, Tegenfeldt JO and Hook F (2006) Improving the instrumental resolution of sensors based on localized surface plasmon resonance. Anal. Chem. 78:4416-4423. DOI: 10.1021/ac0601967 ER.

[6] Lin T and Chung M (2009) Detection of cadmium by a fiber-optic biosensor based on localized surface plasmon resonance. Biosens. Bioelectron. 24:1213-1218. DOI: 10.1016/j.bios.2008.07.013.

[7] Haes AJ, Chang L, Klein WL and Van Duyne RP (2005) Detection of a biomarker for Alzheimer's disease from synthetic and clinical samples using a nanoscale optical biosensor. J. Am. Chem. Soc. 127:2264-2271. DOI: 10.1021/ja044087q

[8] Shen XW, Huang CZ and Li YF (2007) Localized surface plasmon resonance sensing detection of glucose in the serum samples of diabetes sufferers based on the redox reaction of chlorauric acid. Talanta 72:1432-1437. DOI: 10.1016/j.talanta.2007.01.066 ER.

[9] Endo T, Kerman K, Nagatani N, Takamura Y and Tamiya E (2005) Label-free detection of peptide nucleic acid-DNA hybridization using localized surface plasmon resonance based optical biosensor. Anal. Chem. 77:6976-6984. DOI: 10.1021/ac0513459 ER.

[10] Guo LH, Chen GN and Kim DH (2010) Three-Dimensionally Assembled Gold Nanostructures for Plasmonic Biosensors. Anal. Chem. 82:5147-5153. DOI:10.1021/ac100346z ER.

[11] Endo T, Takizawa H, Yanagida Y, Hatsuzawa T and Tamiya E (2008) Construction of a Biosensor Operating on the Combined Principles of Electrochemical Analysis and Localized Surface Plasmon Resonance for Multiple Detection of Antigen-Antibody and Enzymatic Reactions on the Single Biosensor. Sensors and Materials 20:255-265.

[12] Endo T, Kerman K, Nagatani N, Hiepa HM, Kim DK, Yonezawa Y, Nakano K and Tamiya E (2006) Multiple label-free detection of antigen-antibody reaction using localized surface plasmon resonance-based core-shell structured nanoparticle layer nanochip. Anal.Chem. 78:6465-6475. DOI: 10.1021/ac0608321 ER.

[13] Schultz S, Smith DR, Mock JJ and Schultz DA (2000) Single-target molecule detection with nonbleaching multicolor optical immunolabels. Proc. Natl. Acad. Sci. U.S.A. 97:996-1001.

[14] Etchegoin PG and Le Ru EC (2008) A perspective on single molecule SERS: current status and future challenges. Phys. Chem. Chem. Phys. 10:6079-6089. DOI:10.1039/b809196j.

[15] Lakowicz JR, Geddes CD, Gryczynski I, Malicka J, Gryczynski Z, Aslan K, Lukomska J, Matveeva E, Zhang JA, Badugu R and Huang J (2004) Advances in surface-enhanced fluorescence. J. Fluoresc. 14:425-441.

[16] Moskovits M (1985) Surface-Enhanced Spectroscopy. Rev. Mod. Phys. 57:783-826.

[17] Campion A and Kambhampati P (1998) Surface-enhanced Raman scattering. Chem. Soc. Rev. 27:241-250. DOI:10.1039/a827241z

[18] Jensen TR, Schatz GC and Van Duyne RP (1999) Nanosphere lithography: Surface plasmon resonance spectrum of a periodic array of silver nanoparticles by ultraviolet-visible extinction spectroscopy and electrodynamic modeling. J. Phys. Chem. B 103:2394-2401. DOI**:** 10.1021/jp984406y

[19] Gunnarsson L, Rindzevicius T, Prikulis J, Kasemo B, Kall M, Zou SL and Schatz GC (2005) Confined plasmons in nanofabricated single silver particle pairs: Experimental observations of strong interparticle interactions. J Phys Chem B 109:1079-1087. DOI: 10.1021/jp049084e

[20] Nikoobakht B and El-Sayed MA (2003) Preparation and growth mechanism of gold nanorods (NRs) using seed-mediated growth method. Chemistry of Materials 15:1957-1962. DOI: 10.1021/cm020732l ER.

[21] Ditlbacher H, Hohenau A, Wagner D, Kreibig U, Rogers M, Hofer F, Aussenegg FR and Krenn JR (2005) Silver nanowires as surface plasmon resonators. Physical Review Letters DOI:10.1103/PhysRevLett.95.257403 ER.

[22] Rost MJ (2007) In situ real-time observation of thin film deposition: Roughening, zeno effect, grain boundary crossing barrier, and steering. Physical Review Letters DOI: 10.1103/PhysRevLett.99.266101 ER.

[23] Rost MJ, Quist DA and Frenken JWM (2003) Grains, growth, and grooving. Phys.Rev.Lett. 91, 026101 doi: 10.1103/PhysRevLett.91.026101 ER.

[24] Aspnes DE, Kinsbron E and Bacon DD (1980) Optical-Properties of Au - Sample Effects. Physical Review B 21:3290-3299.

[25] Stranahan SM and Willets KA (2010) Super-resolution Optical Imaging of Single-Molecule SERS Hot Spots. Nano Letters 10:3777-3784. DOI:10.1021/nl102559d

[26] Huang WY, Qian W, El-Sayed MA, Ding Y and Wang ZL (2007) Effect of the lattice crystallinity on the electron-phonon relaxation rates in gold nanoparticles. Journal of Physical Chemistry C 111:10751-10757. DOI:10.1021/jp0738917

[27] Rodriguez-Fernandez J, Funston AM, Perez-Juste J, Alvarez-Puebla RA, Liz-Marzan LM and Mulvaney P (2009) The effect of surface roughness on the plasmonic response of individual sub-micron gold spheres. Physical Chemistry Chemical Physics 11:5909-5914. DOI:10.1039/b905200n

[28] Pecharroman C, Perez-Juste J, Mata-Osoro G, Liz-Marzan LM and Mulvaney P (2008) Redshift of surface plasmon modes of small gold rods due to their atomic roughness and end-cap geometry. Physical Review B, 77, 035418. DOI: 10.1103/PhysRevB.77.035418.

[29] Trugler A, Tinguely JC, Krenn JR, Hohenau A and Hohenester U (2011) Influence of surface roughness on the optical properties of plasmonic nanoparticles. Physical Review B 83, 081412(R). DOI: 10.1103/PhysRevB.83.081412.

[30] Ohring M (2002) Materials science of thin films. Academic Press, San Diego.

[31] Thompson CV (2000) Structure evolution during processing of polycrystalline films. Annu. Rev. Mater. Sci. 30:159-190 doi: 10.1146/annurev.matsci.30.1.159

[32] Chen K, Drachev VP, Borneman JD, Kildishev AV and Shalaev VM (2010) Drude relaxation rate in grained gold nanoantennas. Nano Lett. 10:916-922. doi: 10.1021/nl9037246.

[33] Romanyuk VR, Kondratenko OS, Fursenko OV, Lytvyn OS, Zynyo SA, Korchovyi AA and Dmitruk NL (2008) Thermally induced changes in thin gold films detected by polaritonic ellipsometry. Mater. Sci. Eng., B - 149:285-291. doi: 10.1016/j.mseb.2007.10.019 ER.

[34] ,Bohren C and Huffman D (1982) Absorption and scattering of light by small particles. New York: Wiley

[35] Kreibig U, Vollmer M (1995) Optical Properties of metal clusters. New York: Springer.

[36] Kelly KL, Coronado E, Zhao LL and Schatz GC (2003) The optical properties of metal nanoparticles: the influence of size, shape and dielectric environment. The Journal of Physical Chemistry B, 107, 668-677. DOI: 10.1021/jp026731y

[37] Le Ru EC, Etchegoin PG, Grand J, Félidj N, Aubard J, Lévi G, Hohenau A and Krenn JR (2008) Surface enhanced Raman spectroscopy on nanolithography-prepared substrates. Curr. Appl Phys. 8:467-470. doi: 10.1016/j.cap.2007.10.073.

[38] Felidj N, Aubard J, Levi G, Krenn JR, Hohenau A, Schider G, Leitner A and Aussenegg FR (2003) Optimized surface-enhanced Raman scattering on gold nanoparticle arrays. Appl. Phys. Lett. 82:3095-3097. doi: 10.1063/1.1571979.

[39] Le Ru EC, Etchegoin PG, Grand J, Felidj N, Aubard J and Levi G (2007) Mechanisms of spectral profile modification in surface-enhanced fluorescence. J. Phys. Chem.C 111:16076-16079. doi: 10.1021/jp076003g.

# Supporting Information

**Atomic Force Microscopy**

Atomic force microscopy (AFM) measurements were performed using a Digital Instruments Multimode with a Nanoscope IIIa controller. Fig. S1 presents images of gold thin films before (left) and after annealing at 200°C for 20minutes (right) as referred to in section 2.2. The diagonal lines visible at the images are related to the polishing process of the quartz substrates. The root mean square (rms) roughness was deduced by analyzing the whole area of 25µm², excluding the white spots which we assume to be dust particles.

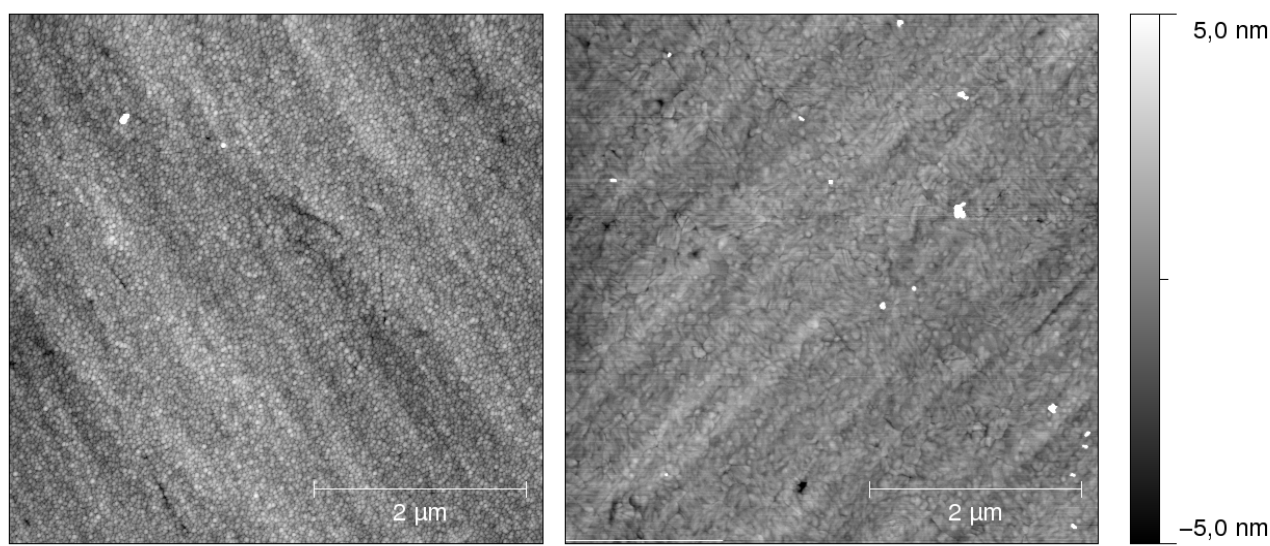

**Fig. S1:** AFM scans of a 40nm thick gold thin film on quartz substrate, before (left) and after (right) annealing at 200°C for 20 minutes.

Fig. S2 shows AFM images of an array of electron-beam lithography (EBL) fabricated gold nanoparticles before (left) and after (right) annealing at 200°C for 5 minutes as described in section 3.1. The rms roughness of the particles was deduced by restricting the analysis to the particle surfaces and applying 3-point leveling and polynomial background correction for sample tilt. Mask-setting is visualized at the inset between the scans.

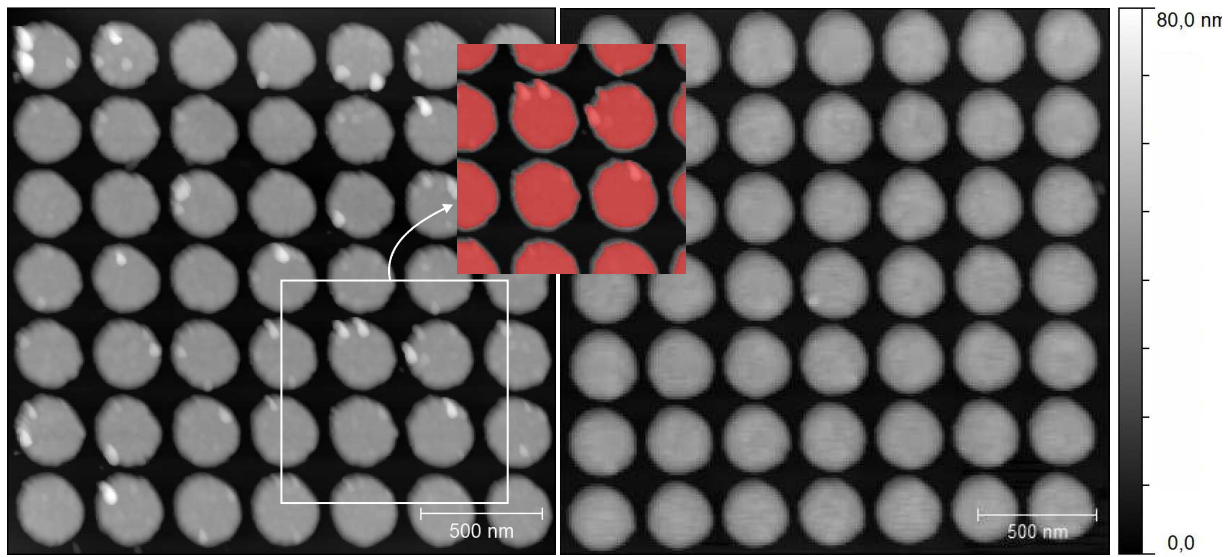

**Fig. S2:** AFM scans of EBL-produced gold nanodiscs before (left) and after annealing (right) at 200°C for 5 minutes. The inset between the scans demonstrates how the particle area (in red) was chosen for rms roughness determination.

**Ellipsometry data**

The dielectric function of a 40nm high gold film was obtained out of ellipsometry measurements (Woolan V.A.S.E. system). Fig.S3 displays the reduction in the real (left) and imaginary (right) terms, $\varepsilon_1$ and $\varepsilon_2$ respectively, before and after annealing at 200°C for 5 minutes compared to literature data by Johnson and Christy [Ref1].

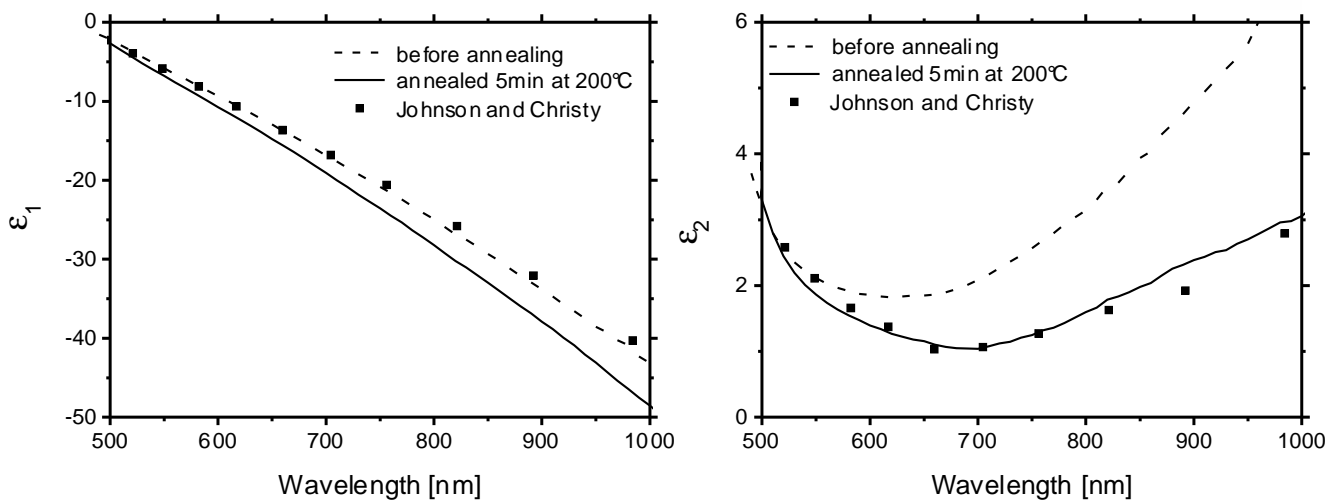

**Fig. S3:** Real (left) and imaginary (right) terms of the dielectric function of a 40nm gold film before (solid line) and after (dashed line) annealing at 200°C for 5 minutes, compared with values provided by literature [Johnson PB and Christy RW (1972), Optical constants of the noble metals. Phys. Rev. B 6:4370-4379, doi: 10.1103/PhysRevB.6.4370]